\documentclass{article}
\usepackage{spconf,amsmath,graphicx}

\usepackage{enumitem}
\setlist{nosep, leftmargin=14pt}

\usepackage{mwe} 

\usepackage{xcolor}

\usepackage{float}
\usepackage[justification=centering]{caption}
\usepackage{hyperref}
\usepackage{comment} 

\usepackage{amssymb}

\title{Brain subtle anomaly detection based on Auto-Encoders latent space analysis: Application to  \textit{de novo} Parkinson patients}

\name{
\begin{tabular}{ccc}
Nicolas Pinon$^{1}$ & Geoffroy Oudoumanessah\sthanks{Has equally contributed as the first author.}$^{1, 2, 3}$ & Robin Trombetta$^{1}$ \\
Michel Dojat$^{2}$  & Florence Forbes$^{3}$ & Carole Lartizien$^{1}$
\end{tabular}}

\address{\normalsize $^{1}$ Univ. Lyon, CNRS UMR 5220, Inserm U1294, INSA Lyon, UCBL, CREATIS, France \\
\normalsize $^{2}$ Univ. Grenoble Alpes, Inserm U1216, CHU Grenoble Alpes, Institut des Neurosciences, France \\
\normalsize $^{3}$ Univ. Grenoble Alpes, Inria, CNRS, Grenoble INP, LJK, France}

\begin{document}

%
\maketitle

\begin{abstract}
Neural network-based  anomaly detection  remains challenging in clinical applications with little or no supervised information and subtle anomalies such as  hardly visible brain lesions. Among unsupervised methods, patch-based auto-encoders with their efficient representation power provided by their latent space,  have shown good results for visible lesion detection. However, the commonly used reconstruction error criterion may  limit their performance when facing less obvious lesions. In this work, we design two alternative detection criteria. They are derived from multivariate analysis and can more directly capture information from latent space representations. 
Their performance  compares favorably  with  two additional supervised learning methods, on a difficult \textit{de novo} Parkinson Disease (PD) classification task.
\end{abstract}


\vspace{-15px}  
\section{Introduction}
\label{sec:intro}
Most recent success of deep supervised learning, in the context of medical image analysis, critically depends on the availability of large sets of annotated images. The performance of supervised learning methods, on tasks such as anomaly detection, is then limited when the studied pathology is rare or when a fine expert annotation is required. A typical example is that of \textit{de novo} (just diagnosed) Parkinson Disease (PD) patients, for which  brain structural abnormalities are subtle and hardly visible
in standard T1w or diffusion MR images.
A natural alternative to supervised methods is \textit{outlier detection} or \textit{Unsupervised Anomaly Detection} (UAD).
This formalism requires only the manual identification of "normal" data to construct a tractable model of normality, while \textit{outliers} are then automatically detected as samples deviating from this normal model.
Different categories of UAD methods have been applied to medical image segmentation or detection tasks. They mainly differ in the features used to learn the normal model and the score computed to assess the distance to this model, which in brain anomaly detection is typically assessed at the voxel level. 
Illustrations include several auto-encoders (AE) architectures that have been compared 
in \cite{baur_autoencoders_2020}. These AE models 
are trained to perform a  ``pretext'' task on normal images consisting in the reconstruction of these images. For an arbitrary image, voxel-wise anomaly scores are then computed  as the reconstruction errors, \textit{i.e.} the differences between the image voxels and the reconstructed ones.  Such errors are expected to be much larger for unseen voxels from patient images, provided the chosen architecture has initially well captured the normal subjects main features.
To further investigate the importance of the normal model construction, building on this standard deep UAD formalism, we recently compared different auto-encoders architectures for the detection of subtle anomalies in the diffusion parametric maps of \textit{de novo} PD patients \cite{mlcn2021}. This comparison included an auto-encoder (AE), taking 2D transverse slices as input, and the adaptation of a patch-based siamese auto-encoder (SAE) proposed in \cite{alaverdyan_regularized_2020}. Our results 
demonstrated encouraging performance with the SAE model slightly outperforming the AE, thus indicating that patches may indeed be advantageous, in particular for their ability to capture local spatial neighborhood information around each voxel. 
However, as regards the detection score, the study also 
confirmed recent observations outlining the limitations of the reconstruction error scores for the detection of very subtle abnormalities \cite{meissen2022_pitfalls}.
In this work, we propose to investigate other detection procedures combining  1) enhanced normal models and 2) scoring rules derived from multivariate statistics. 
Following the approach reported in \cite{alaverdyan_regularized_2020}, we consider a patch-based approach but 
propose to perform the detection step in the latent space of the auto-encoder. 
More specifically, latent space representations of the normal images are extracted from  the patch-based SAE of \cite{mlcn2021}, and then used as features to build a normal model. Two types of models are considered, a non parametric discriminative one class support vector machine (OC-SVM) \cite{ElAzami_PlosOne2016} and a parametric generative mixture model \cite{arnaud_fully_2018} (see Figure \ref{SAE_diagram} and next section for details). So doing, the hope is to combine the representation power of patch-based AE networks to extract relevant and subtle features, with the efficiency of multivariate statistical models. 
These two combinations are then compared to a baseline UAD model based on the reconstruction error and to two standard \textit{supervised} CNN, namely 3D ResNet and DenseNet.  \\

\vspace{-15px}  

\section{UAD pipeline}
\label{sec:methods}
\begin{figure*}
    \centering
    \includegraphics[scale=0.092]{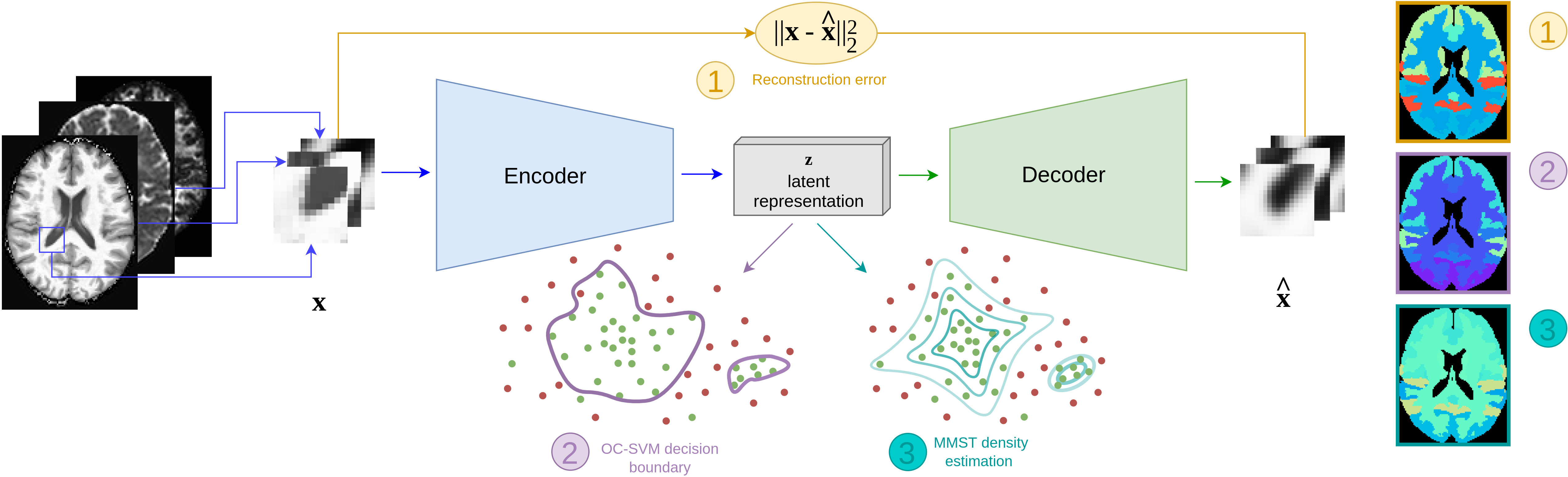}
    \caption{\small The trained encoder extracts latent representation $\mathbf{z}$ of patches, used by 1) a decoder to compute reconstruction error in the image space 2) OC-SVM and 3) $\mathcal{MMST}$ to perform outlier detection in the latent space. Anomaly maps representing the percentage of abnormal voxels per brain structures are shown on the right, warm colors corresponding to the highest percentages.}
    \label{SAE_diagram}
\end{figure*}

The proposed framework for unsupervised brain anomaly detection is depicted on Figure \ref{SAE_diagram}. The central AE is first trained to learn the representation space of normal samples and reconstruct pseudo-normal images. The standard setting consists of computing 
\textit{reconstruction error} maps (as the difference between the input and output images) on which anomalous unseen regions  are expected to exhibit poor reconstructions or equivalently high anomaly scores. 
In this work, we also investigate
two other outlier detection rules based respectively on a \textit{generative} and a \textit{discriminative} model designed to capture information from the AE latent space. 
\vspace{-10px}  
\subsection{Latent space feature extraction}
To construct an efficient normal model, we consider a patch-based network to enrich the latent space with local information at the voxel level. 
Leveraging the architecture proposed in \cite{alaverdyan_regularized_2020}, we use a SAE \cite{lecun_siamese} composed of two replica of an auto-encoder sharing the same weights, associated to the following loss term :

$L_{SAE}(\mathbf{x_1}, \mathbf{x_2}) =  \sum_{t=1}^{2} { {||\mathbf{x_{t}} -\mathbf{\hat{x}_{t}}||_2^2}} - \alpha \cdot  {cos(\mathbf{z_{1}}, \mathbf{z_{2}})}$

\noindent which balances two objectives 1) decoding the representation $\mathbf{z}$ learned from the encoder fed with patches $\mathbf{x}$ into a reconstruction $\mathbf{\hat{x}}$ that is close  from the original patch $\mathbf{x}$ and 2) having close (in the sens of the cosine similarity) $\mathbf{z}$ for similar patches\footnote{In the case of learning on brain MR images,``similar'' patches means that the patches are located in the same place in the brain, which is possible because all MRI's are registered to a common atlas beforehand.}. 


\vspace{-10px}  

\subsection{Outlier detection in the latent space}

As an alternative to the reconstruction error $|| \mathbf{x} - \mathbf{\hat{x}}||_2^2$ between a 
 patch ${\mathbf{x}}$ and its reconstruction $\mathbf{\hat{x}}$,
 we present below two outlier detection procedures built from a  collection of normal patch representations $(\mathbf{z}_i)_{1 \leq i \leq n}$ to account for normality in the latent space.



\noindent {\bf A discriminative approach: One-Class SVM.} \hfill \break
\noindent The goal of the OC-SVM \cite{scholkopf_support_1999} is to construct a decision function $f$, positive on the estimated support of the distribution of normal samples $\mathbf{z}_i$, negative elsewhere and null on the frontier. 
The training samples from the normal class are first mapped to a higher dimensional space via a feature map $\boldsymbol{\phi}(\cdot)$ associated with a kernel $k$ such that $k(\mathbf{z}_i,\mathbf{z}_j)$ = $\boldsymbol{\phi}(\mathbf{z}_i) \cdot \boldsymbol{\phi}(\mathbf{z}_j)$.
As the problem is linear in this redescription space, the parameters $\boldsymbol{w}$ and $\rho$ of the hyperplane $\boldsymbol{w} \cdot \boldsymbol{\phi}(\mathbf{z}) - \rho = 0$ are obtained by solving a convex optimization problem aiming at maximizing the distance of the hyperplane from the origin.

\noindent The decision function can then be expressed as $f(\mathbf{z}) = \boldsymbol{w} \cdot \boldsymbol{\phi}(\mathbf{z}) -\rho$.
In a typical scenario, samples with negatives scores of $f$ would be considered outliers. During inference, $\mathbf{z}$ extracted from patches can be evaluated by the decision function to get an anomaly score corresponding to their distance to the hyperplane. This anomaly score, 
attributed to the central voxel of each patch, then provides an anomaly score map for the whole image. An ensemble of OC-SVM scores, trained on different $\mathbf{z_i}$ is used to provide a more robust anomaly map. \hfill \break

\vspace{-10px}  

\noindent {\bf A generative approach: multivariate mixtures.} \hfill \break
While OC-SVM  estimates only the support of the normal model, the goal here is to estimate the full normal distribution.
To this end, we use a mixture model distribution $p$, denoted by $\mathcal{MMST}$, whose individual components are  multiple scale t-distributions ($\mathcal{MST}$):

$p(\mathbf{z}; \boldsymbol{\Theta}) = \sum _{k=1}^{K} \pi_k \mathcal{MST}(\mathbf{z}; \boldsymbol{\theta}_k)$

\noindent with $\boldsymbol{\Theta} = (\pi_k, \boldsymbol{\theta}_k)_{1 \leq k \leq K}$, $\pi_k \in [0,1]$ and $\sum\limits_{k=1:K} \pi_k = 1$

\noindent $\mathcal{MST}$ distributions are  generalizations of the multivariate t-distribution that extend its Gaussian scale mixture representation. The standard univariate scale variable is replaced by a $M$-dimensional scale variable  $(W_m)_{1 \leq m \leq M} \in \mathbb{R}^M$ where $M$ denotes the latent space dimension. This allows a richer variety of shapes beyond elliptical distributions. The scale variable $W_m$ for dimension $m$ can be interpreted as accounting for the reliability of this dimension and is typically small when $\mathbf{z}$ is far from the mean parameter. The specific definition can be found in \cite{MST}. 
Given a learning set of $(\mathbf{z}_i)_{1 \leq i \leq n}$, 
the estimation of the model parameter denoted by $\hat{\boldsymbol{\Theta}}_n$ is theoretically feasible using a standard expectation-maximization (EM) algorithm  but is too time and memory costly in practice when the amount of data is large. In this work, we therefore resort to an  \textit{online}
version of EM \cite{OEM} that we derived for our $\mathcal{MMST}$  model as detailed  in \cite{rapportG}.
Finally given a latent representation of a patch $\mathbf{z}$, we can use the scale variables to derive a measure of proximity $f$ to the learned normal model:
 $f(\mathbf{z}) = \max_{1 \leq m \leq M} \Bar{w}_m^{\mathbf{z}}$, 
 with $\Bar{w}_m^{\mathbf{z}} = \mathbb{E}[W_m | \mathbf{z}; \hat{\boldsymbol{\Theta}}_n]$, where the expectation is computed for the learned $\mathcal{MMST}$ model and is typically larger when at least one dimension of $\mathbf{z}$ is well explained by the model. 
This measure of proximity, available for each voxel,  provides in turn an anomaly score map for the whole image.

\vspace{-10px}  

\subsection{Post-processing of the anomaly maps}
\label{sec:postpro}
\begin{figure*}[!ht]
    \centering
    \includegraphics[scale=0.057]{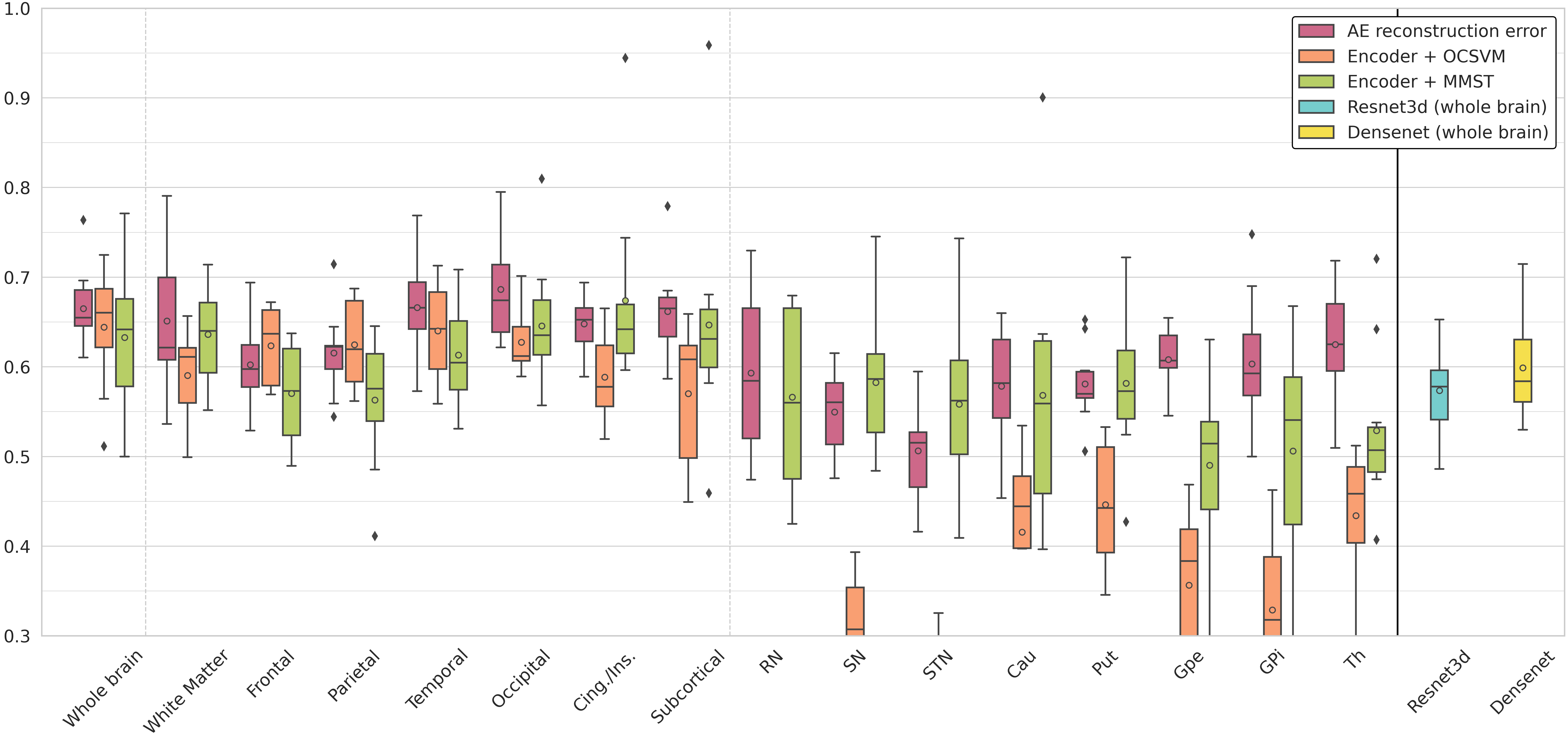}
    \caption{\small \textit{g-mean score} of the 3 UAD and 2 CNN models. For UAD models, we consider anomaly \% on the whole brain and per region, including the 8 subcortical structures from the MNI PD25 atlas: substantia nigra  (SN), red nucleus (RN), subthalamic nucleus (STN), globus pallidus interna and externa (GPi, GPe), thalamus, putamen and caudate nucleus.}
    \label{boxplot}
\end{figure*}

A threshold value (the \textit{abnormality threshold}) set to an extreme quantile (eg. in the range of [90\%, 100\%[) of the anomaly scores distribution in the normal train samples was derived for each method (reconstruction error, encoder + OC-SVM and encoder + $\mathcal{MMST}$) and applied to the test patient and test control dataset. The resulting binary anomaly maps can serve to identify suspect regions. To help evaluate the localization of these anomalies, two atlases were considered and fused: the Neuromorphometrics atlas \cite{neuromorphometrics} which segments the brain into 8 macro-regions and the MNI PD25 atlas \cite{Xiao2015} which is specifically  designed  for  PD  patients exploration and delineates 8 relevant subcortical structures (see Fig \ref{boxplot}). The percentage of anomalous voxels was computed for each of these regions of interest leading to region-wise anomaly maps as depicted on the right of Figure \ref{SAE_diagram}.

\vspace{-10px}
\section{Experiments}

\subsection{Data description and splitting}

\label{sec:data}
T1-weighted and DTI MR scans from 54 healthy controls and 124 \textit{de novo} PD patients were extracted from the PPMI database \cite{ppmi}. All retrieved images were acquired with the same MR scanner model (3T Siemens Trio Tim).
Mean diffusivity (MD) and fractional anisotropy (FA) maps were computed from DTI using MRtrix3.0. 
All maps $X$  (T1w, FA, MD) were normalized in intensity with \hfill \break $X_{\text{norm}} = \frac{X - 1\%quantile(\chi)}{99\%quantile(\chi) - 1\%quantile(\chi)}$ with $\chi$ being the intensity distribution of train controls images of one modality. \hfill \break
All maps were non-linearly registered onto the MNI atlas resulting in images of dimension $121 \times 145 \times 121$ with a voxel size of $1.5 \times 1.5 \times 1.5\  \text{mm}^3$. As for the cross-validation, healthy controls dataset was divided into 10 folds following a bootstrap procedure \cite{Poldrack2019}, leading to each fold containing $[\![$39, 41$]\!]$ train controls and $[\![$13, 15$]\!]$ test controls. The same procedure was performed with PD patients, leading to each fold containing $[\![$36, 40$]\!]$ train patients and $[\![$82, 86$]\!]$ test patients. Special care was put into balancing the age and sex distribution of each fold.
\vspace{-10px}
\subsection{Hyperparameters of the UAD pipeline}

The encoder was composed of 4 convolutionnal blocks with kernel size $(5, 5)$, $(3, 3)$, $(3, 3)$ and $(3, 3)$, with strides respectively $(1, 1)$, $(1, 1)$, $(3, 3)$ and $(1, 1)$, number of filters respectively $3$, $4$, $12$ and $16$, no padding and GeLu activation. Each block was followed by a batch normalization block. 
The decoder was the symmetric counterpart of the encoder. The input of the encoder consisted of the patches of each of the 3 modalities combined as channels. The SAE model was trained with $[\![$975000, 1025000$]\!]$ patches of size  15$\times$15$\times$3 (25 000 patches per subject).
We used Adam optimizer \cite{kingma_adam_2017} for 20 epochs, with default hyperparameters, best model selection based on validation loss and training batch size of 1000. 
An ensemble of five OC-SVM were trained, each with 500 $\mathbf{z}_i$ samples extracted from 500 random brain localizations from the train set and the mean of the 5 decision functions was used as the final anomaly score (note that this differs from \cite{alaverdyan_regularized_2020} where one OC-SVM is trained per voxel). We used $\nu = 0.03$ and a Gaussian kernel whose hyperparameter $\frac{1}{\gamma}$ was set to the product of variance and dimension of the $\mathbf{z}_i$. For $\mathcal{MMST}$, we used $K = 9$.  We set the \textit{abnormality threshold} defined in section \ref{sec:postpro} to 98\% (experiments have shown that the choice of this threshold has little influence on the final performance).



\vspace{-10px}
\subsection{Performance evaluation of the UAD models}
Performance of the three methods was evaluated as in \cite{arnaud_fully_2018, mlcn2021}. The percentage of abnormal voxels in the whole brain or per region of interest derived from  the post-processing of the anomaly score maps (see section \ref{sec:postpro}) 
was employed to classify the test controls and test patients as healthy or pathological (PD). 
By varying a threshold on this metric,
we can draw a ROC curve from the test population, and derive the best-achievable \textit{g-mean score}  defined as $\sqrt{\text{Sensitivity} \times \text{Specificity}}$. \textit{g-mean score} is used as a performance metric to compare the different classification models. In the absence of reference annotations of the brain structures affected by the pathology, this pretext classification task allows to indirectly evaluate if the anomalies detected by the UAD models are characteristic of the pathology. It was computed either considering the percentage of anomalies in the whole brain, or in each of the regions of interest of the Neuromorphometric and MNI PD25 atlases.

\vspace{-10px}  
\subsection{Comparison with supervised approaches}
\label{supervised}
We compared classification performance of the three UAD models (reconstruction error, encoder + OC-SVM and encoder + $\mathcal{MMST}$) to that of two standard supervised 3D convolutional networks: 3D ResNet with 18 layers \cite{resnet3d} and DenseNet-264 \cite{densenet}. Each of these 2 CNN took as input the whole 3D T1w, MD and FA brain images combined as channels. A dense layer was added at the end of each network in order to have a one-dimensional output for classification. For each fold, the models were trained on 75\% of train controls and train patients, the remaining 25\% being kept for validation. Training was performed with Adam optimizer \cite{kingma_adam_2017} for 300 epochs with default hyperparameters and a batch size of 8. Note that the train patients described in section \ref{sec:data} were only used for training of these two supervised networks. These two models were evaluated on the same test patient dataset as used for the UAD models thus enabling a fair comparison.

\vspace{-10px}  

\section{Results}
\label{sec:results}

The \textit{g-mean score} of each method is reported in Figure \ref{boxplot}. \break
We notice that the 3 UAD models achieve a median \textit{g-mean score} around 0.65 on the whole brain, and in the range [0.6, 0.7] when only considering certain macro-regions (e.g. temporal or occipital lobe). For subcortical structures (e.g. RN or SN), performance drop to the range [0.5, 0.6] and even lower for some methods (especially Encoder + OC-SVM). At this stage of the PD progression, these subcortical structures seem slightly impacted.  
Note that the supervised methods, Resnet3D and Densenet, provide on the whole brain a median \textit{g-mean score} in the range [0.55, 0.6], lower than the UAD models considered in this study. 

\vspace{-10px}
\section{Discussion and conclusion}
\label{ssec:discussion}
Auto-encoders have shown to be a reference method regarding unsupervised anomaly detection  \cite{baur_autoencoders_2020} but have also shown limits when used for very subtle anomalies \cite{meissen2022_pitfalls}. We have investigated whether an analysis of the latent space could improve these performances compared to a classical reconstruction error approach. We used two methods based on different paradigms: One-Class SVM (\textit{discriminative}) and Mixture of Multiple scaled t-distributions (\textit{generative}). It is clear from the supervised networks results that the proposed task, discriminating \textit{de novo} PD from controls, is very hard: the supervised methods performances fall below the unsupervised methods ones, validating our approach. 
As seen with the performance of the reconstruction error, we found that the latent space UAD methods are strong competitors but do not surpass the former.
In comparison with \cite{mlcn2021} where only diffusion was used, we report that the addition of T1w images does not improve significantly the performances. \hfill \break
We also demonstrated that using a patch-based encoder, as a feature extractor to feed a $\mathcal{MMST}$ model, gave promising results as it allows capturing some spatial context, which was lacking in \cite{arnaud_fully_2018}. Finally, the discrimination of PD based only on subcortical structures seems not feasible, as reported in \cite{prasuhn_machine_2020} for substantia nigra, at an early stage of the pathology. \hfill \break 
Future work includes investigating whether the combination of reconstruction error and latent space anomaly maps can increase the classification performance. We aim to extract 3D features with the auto-encoders and complete the multi-modal approach by adding T2w and T2$^*$w images as in \cite{sivaranjini_deep_2020}.

\vspace{-10px}  

\section{Acknowledgments}
\label{sec:acknowledgments}
G. Oudoumanessah was financially supported by the AURA region. 
This work was granted access to the HPC resources of IDRIS under the allocation 2022-AD011012813R1 made by GENCI. It was partially funded by French program
“Investissement d’Avenir” run by the Agence Nationale pour
la Recherche (ANR-11-INBS-0006).

\vspace{-10px}  

\section{Compliance with Ethical Standards}
This research study was conducted retrospectively using human subject data made available by the Parkinson
Progression Markers Initiative (PPMI). Ethical approval was not required as confirmed by the license attached with the open access data.


\vspace{-10px}  


\bibliographystyle{IEEEbib}
\bibliography{strings,refs}

\begin{thebibliography}{10}

\bibitem{baur_autoencoders_2020}
C.~Baur, S.~Denner, B.~Wiestler, et~al.,
\newblock ``{Autoencoders for unsupervised anomaly segmentation in brain MR
  images: A comparative study},''
\newblock {\em Medical Image Analysis}, vol. 69, 2021.

\bibitem{mlcn2021}
V.~Muñoz-Ramírez, N.~Pinon, F.~Forbes, et~al.,
\newblock ``Patch vs. {Global} {Image}-{Based} {Unsupervised} {Anomaly}
  {Detection} in {MR} {Brain} {Scans} of {Early} {Parkinsonian} {Patients},''
\newblock in {\em Machine {Learning} in {Clinical} {Neuroimaging}}, Cham, 2021,
  pp. 34--43.

\bibitem{alaverdyan_regularized_2020}
Z.~Alaverdyan, J.~Jung, R.~Bouet, and C.~Lartizien,
\newblock ``Regularized siamese neural network for unsupervised outlier
  detection on brain multiparametric magnetic resonance imaging: {Application}
  to epilepsy lesion screening,''
\newblock {\em Medical Image Analysis}, vol. 60, 2020.

\bibitem{meissen2022_pitfalls}
F.~Meissen, B.~Wiestler, G.~Kaissis, and D.~Rueckert,
\newblock ``On the pitfalls of using the residual as anomaly score,''
\newblock in {\em Medical Imaging with Deep Learning}, 2022.

\bibitem{ElAzami_PlosOne2016}
M.~El~Azami, A.~Hammers, J.~Jung, et~al.,
\newblock ``Detection of lesions underlying intractable epilepsy on t1-weighted
  mri as an outlier detection problem,''
\newblock {\em PLOS ONE}, vol. 11, no. 9, pp. 1--21, 09 2016.

\bibitem{arnaud_fully_2018}
A.~Arnaud, F.~Forbes, N.~Coquery, et~al.,
\newblock ``Fully automatic lesion localization and characterization:
  Application to brain tumors using multiparametric quantitative mri data,''
\newblock {\em IEEE TMI}, vol. 37, no. 7, 2018.

\bibitem{lecun_siamese}
J.~Bromley, I.~Guyon, Y.~LeCun, et~al.,
\newblock ``Signature verification using a "siamese" time delay neural
  network,''
\newblock in {\em NIPS}, 1993, p. 737–744.

\bibitem{scholkopf_support_1999}
B.~Schölkopf, R.~C.Williamson, A.~Smola, et~al.,
\newblock ``Support {Vector} {Method} for {Novelty} {Detection},''
\newblock in {\em Advances in Neural Information Processing Systems}, S.~Solla,
  T.~Leen, and K.~Müller, Eds., 1999, vol.~12.

\bibitem{MST}
F.~Forbes and D.~Wraith,
\newblock ``A new family of multivariate heavy-tailed distributions with
  variable marginal amounts of tailweight: application to robust clustering,''
\newblock {\em Stat. and computing}, vol. 24, no. 6, pp. 971--984, 2014.

\bibitem{OEM}
O.~Capp{\'e} and E.~Moulines,
\newblock ``On-line expectation--maximization algorithm for latent data
  models,''
\newblock {\em JRSS B}, vol. 71, no. 3, pp. 593--613, 2009.

\bibitem{rapportG}
G.~Oudoumanessah, M.~Dojat, and F.~Forbes,
\newblock ``{Unsupervised scalable anomaly detection: application to medical
  imaging},''
\newblock Research report, Oct. 2022,
\newblock \url{https://hal.archives-ouvertes.fr/hal-03824951}.

\bibitem{neuromorphometrics}
R.~Bakker, P.~Tiesinga, and R.~Kotter,
\newblock ``{{T}he {S}calable {B}rain {A}tlas: {I}nstant {W}eb-{B}ased {A}ccess
  to {P}ublic {B}rain {A}tlases and {R}elated {C}ontent},''
\newblock {\em Neuroinformatics}, vol. 13, pp. 353--366, 2015.

\bibitem{Xiao2015}
Y.~Xiao, V.~Fonov, S.~Beriault, et~al.,
\newblock ``{{M}ulti-contrast unbiased {M}{R}{I} atlas of a {P}arkinson's
  disease population},''
\newblock {\em Int J Comput Assist Radiol Surg}, vol. 10, pp. 329--341, 2015.

\bibitem{ppmi}
Kenneth Marek, Sohini Chowdhury, Andrew Siderowf, et~al.,
\newblock ``The parkinson's progression markers initiative (ppmi) -
  establishing a pd biomarker cohort,''
\newblock {\em Annals of Clinical and Translational Neurology}, p. 1460–1477,
  2018.

\bibitem{Poldrack2019}
Russell~A. Poldrack, Grace Huckins, and Gael Varoquaux,
\newblock ``{Establishment of Best Practices for Evidence for Prediction: A
  Review},''
\newblock {\em JAMA Psychiatry}, pp. 534--540, 2019.

\bibitem{kingma_adam_2017}
Diederik~P Kingma and Jimmy Ba,
\newblock ``Adam: A method for stochastic optimization,''
\newblock in {\em ICLR (Poster)}, 2015.

\bibitem{resnet3d}
D.~Tran, H.~Wang, L.~Torresani, et~al.,
\newblock ``A closer look at spatiotemporal convolutions for action
  recognition,''
\newblock in {\em 2018 IEEE CVPR}, 2018, pp. 6450--6459.

\bibitem{densenet}
G.~Huang, Z.~Liu, G.~Pleiss, et~al.,
\newblock ``Convolutional networks with dense connectivity,''
\newblock {\em IEEE PAMI}, 2019.

\bibitem{prasuhn_machine_2020}
J.~Prasuhn, M.~Heldmann, T.~F. Münte, and N.~Brüggemann,
\newblock ``A machine learning-based classification approach on {Parkinson}’s
  disease diffusion tensor imaging datasets,''
\newblock {\em Neurological Research and Practice}, vol. 2, no. 1, pp. 46, Dec.
  2020.

\bibitem{sivaranjini_deep_2020}
S.~Sivaranjini and C.~M. Sujatha,
\newblock ``Deep learning based diagnosis of {Parkinson}’s disease using
  convolutional neural network,''
\newblock {\em Multimedia Tools and Applications}, vol. 79, no. 21-22, pp.
  15467--15479, June 2020.

\end{thebibliography}



@inproceedings{mlcn2021,
	address = {Cham},
	title = {Patch vs. {Global} {Image}-{Based} {Unsupervised} {Anomaly} {Detection} in {MR} {Brain} {Scans} of {Early} {Parkinsonian} {Patients}},
	isbn = {978-3-030-87586-2},
	booktitle = {Machine {Learning} in {Clinical} {Neuroimaging}},
	author = {Muñoz-Ramírez, V. and Pinon, N. and Forbes, F. and Lartizen, C. and Dojat, M.},
	year = {2021},
	pages = {34--43},
}

@article{Alaverdyan2020,
author = {Alaverdyan, Z. and Jung, J. and Bouet, R. and Lartizien, C.},
doi = {10.1016/j.media.2019.101618},
journal = {Medical Image Analysis},
pages = {101618},
title = {{Regularized siamese neural network for unsupervised outlier detection on brain multiparametric magnetic resonance imaging: Application to epilepsy lesion screening}},
volume = {60},
year = {2020}
}

@INPROCEEDINGS{Munoz_isbi,
  author={Muñoz Ramírez, V. and Kmetzsch, V. and Forbes, F. and Dojat, M.},
  booktitle={2020 IEEE 17th International Symposium on Biomedical Imaging (ISBI)}, 
  title={Deep Learning Models to Study the Early Stages of Parkinson's Disease}, 
  year={2020},
  pages={1534-1537},
  doi={10.1109/ISBI45749.2020.9098529}}
  
@article{Munoz_AIM,
	author = {Mu{\~n}oz-Ram{\'\i}rez, Ver{\'o}nica and Kmetzsch, Virgilio and Forbes, Florence and Meoni, Sara and Moro, Elena and Dojat, Michel},
	title = {Subtle anomaly detection in MRI brain scans: Application to biomarkers extraction in patients with de novo Parkinson{\textquoteright}s disease},
	year = {2021},
	doi = {10.1101/2021.06.03.21258269},
	publisher = {Cold Spring Harbor Laboratory Press},
	eprint = {https://www.medrxiv.org/content/early/2021/06/05/2021.06.03.21258269.full.pdf},
	journal = {medRxiv}
}

@ARTICLE{arnaud_fully_2018,
  author={Arnaud, A. and Forbes, F. and Coquery, N. and Collomb, N. and Lemasson, B. and Barbier, E. L.},
  journal={IEEE TMI}, 
  title={Fully Automatic Lesion Localization and Characterization: Application to Brain Tumors Using Multiparametric Quantitative MRI Data}, 
  year={2018},
  volume={37},
  number={7},
  doi={10.1109/TMI.2018.2794918}}


@article{alaverdyan_regularized_2020,
	title = {Regularized siamese neural network for unsupervised outlier detection on brain multiparametric magnetic resonance imaging: {Application} to epilepsy lesion screening},
	volume = {60},
	issn = {13618415},
	shorttitle = {Regularized siamese neural network for unsupervised outlier detection on brain multiparametric magnetic resonance imaging},
	url = {https://linkinghub.elsevier.com/retrieve/pii/S1361841519301562},
	doi = {10.1016/j.media.2019.101618},
	language = {en},
	urldate = {2020-11-02},
	journal = {Medical Image Analysis},
	author = {Alaverdyan, Z. and Jung, J. and Bouet, R. and Lartizien, C.},
	year = {2020},
	keywords = {\_lu, siamese networks}
}

@article{decompt,
  title={Gaussian parsimonious clustering models},
  author={Celeux, G. and Govaert, G.},
  journal={Pattern recognition},
  volume={28},
  number={5},
  pages={781--793},
  year={1995},
  publisher={Elsevier}
}

@book{EM,
  title={The EM algorithm and extensions},
  author={McLachlan, G. J. and Krishnan, T.},
  year={2007},
  publisher={John Wiley \& Sons},
  note = {John Wiley \& Sons}
}

@article{MST,
  title={A new family of multivariate heavy-tailed distributions with variable marginal amounts of tailweight: application to robust clustering},
  author={Forbes, F. and Wraith, D.},
  journal={Stat. and computing},
  volume={24},
  number={6},
  pages={971--984},
  year={2014},
  publisher={Springer}
}

@article{OEM,
  title={On-line expectation--maximization algorithm for latent data models},
  author={Capp{\'e}, O. and Moulines, E.},
  journal={JRSS B},
  volume={71},
  number={3},
  pages={593--613},
  year={2009},
  publisher={Wiley Online Library}
}

@techreport{rapportG,
  TITLE = {{Unsupervised scalable anomaly detection: application to medical imaging}},
  AUTHOR = {Oudoumanessah, G. and Dojat, M. and Forbes, F.},
  URL = {https://hal.archives-ouvertes.fr/hal-03824951},
  TYPE = {Research Report},
  YEAR = {2022},
  MONTH = Oct,
  PDF = {https://hal.archives-ouvertes.fr/hal-03824951/file/r.pdf},
  HAL_ID = {hal-03824951},
  HAL_VERSION = {v1},
  note={\url{https://hal.archives-ouvertes.fr/hal-03824951}}
}



@book{scholkopf_learning_2002,
	series = {Adaptive computation and machine learning},
	title = {Learning with kernels: support vector machines, regularization, optimization, and beyond},
	isbn = {978-0-262-19475-4},
	shorttitle = {Learning with kernels},
	language = {en},
	publisher = {MIT Press},
	author = {Schölkopf, B. and Smola, A. J.},
	year = {2002},
	keywords = {Kernel functions, Support vector machines},
}


@inproceedings{scholkopf_support_1999,
	title = {Support {Vector} {Method} for {Novelty} {Detection}},
	booktitle = {Advances in Neural Information Processing Systems},
		volume = {12},
	url = {https://proceedings.neurips.cc/paper/1999/file/8725fb777f25776ffa9076e44fcfd776-Paper.pdf},
	author = {B. Schölkopf and R. C.Williamson, and A. Smola, and Shawe-Taylor, John and Platt, John},
	editor = {Solla, S. and Leen, T. and Müller, K.},
	year = {1999},
}


@article{kingma_auto-encoding_2014,
	title = {Auto-{Encoding} {Variational} {Bayes}},
	url = {},
	language = {en},
	urldate = {2020-11-24},
	journal = {arXiv:1312.6114 [cs, stat]},
	author = {Kingma, Diederik P. and Welling, Max},
	month = may,
	year = {2014}
}

@inproceedings{kingma_adam_2017,
  title={Adam: A Method for Stochastic Optimization},
  author={Kingma, Diederik P and Ba, Jimmy},
  booktitle={ICLR (Poster)},
  year={2015}
}

@article{ioffe_batch_2015,
	title = {Batch {Normalization}: {Accelerating} {Deep} {Network} {Training} by {Reducing} {Internal} {Covariate} {Shift}},
	journal = {arXiv:1502.03167 [cs]},
	author = {Ioffe, Sergey and Szegedy, Christian},
	month = mar,
	year = {2015}
}

@article{baur_autoencoders_2020,
author = {Baur, C. and Denner, S. and Wiestler, B. and Navab, N. and Albarqouni, S.},
doi = {https://doi.org/10.1016/j.media.2020.101952},
journal = {Medical Image Analysis},
title = {{Autoencoders for unsupervised anomaly segmentation in brain MR images: A comparative study}},
volume = {69},
year = {2021}
}

@inproceedings{lecun_siamese,
author = {Bromley, J. and Guyon, I. and LeCun, Y. and S\"{a}ckinger, E. and Shah, R.},
title = {Signature Verification Using a "Siamese" Time Delay Neural Network},
year = {1993},
abstract = {This paper describes an algorithm for verification of signatures written on a pen-input tablet. The algorithm is based on a novel, artificial neural network, called a "Siamese" neural network. This network consists of two identical sub-networks joined at their outputs. During training the two sub-networks extract features from two signatures, while the joining neuron measures the distance between the two feature vectors. Verification consists of comparing an extracted feature vector with a stored feature vector for the signer. Signatures closer to this stored representation than a chosen threshold are accepted, all other signatures are rejected as forgeries.},
booktitle = {NIPS},
pages = {737–744}
}

@inproceedings{meissen2022_pitfalls,
title={On the Pitfalls of Using the Residual as Anomaly Score},
author={ Meissen, F.  and  Wiestler, B. and  Kaissis, G. and  Rueckert, D.},
booktitle={Medical Imaging with Deep Learning},
year={2022},
url={https://openreview.net/forum?id=ZsoHLeupa1D}
}


@INPROCEEDINGS{resnet3d,  
author={Tran, D. and Wang, H. and Torresani, L. and Ray, J. and LeCun, Y. and Paluri, M.},  booktitle={2018 IEEE CVPR},   
title={A Closer Look at Spatiotemporal Convolutions for Action Recognition},   
year={2018}, pages={6450-6459}}

@article{densenet,
 title={Convolutional Networks with Dense Connectivity},
 author={Huang, G. and Liu, Z. and Pleiss, G. and Van Der Maaten, L. and Weinberger, K.},
 journal={IEEE PAMI},
 year={2019}
 }


@article{ppmi,
author = {Marek, Kenneth and Chowdhury, Sohini and Siderowf, Andrew and Lasch, Shirley and Coffey, Christopher and Caspell-Garcia, Chelsea and Simuni, Tanya and Jennings, Danna and M. Tanner, Caroline and Q. Trojanowski, John and Shaw, Leslie and Seibyl, John and Schuff, Norbert and Singleton, Andrew and Kieburtz, Karl and W. Toga, Arthur and Mollenhauer, Brit and Galasko, Doug and M. Chahine, Lana and Parkinson's Progression Markers Initiative},
year = {2018},
pages = {1460–1477},
title = {The Parkinson's progression markers initiative (PPMI) - establishing a PD biomarker cohort},
journal = {Annals of Clinical and Translational Neurology}
}

@article{Zhao2010,
author = {Zhao, Ying Jiao and Wee, Hwee Lin and Chan, Yiong-Huak and Seah, Soo Hoon and Au, Wing Lok and Lau, Puay Ngoh and Pica, Emmanuel Camara and Li, Shu Chuen and Luo, Nan and Tan, Louis C.S.},
doi = {10.1002/mds.22875},
journal = {Movement Disorders},
month = {apr},
number = {6},
pages = {710--716},
publisher = {Wiley-Blackwell},
title = {{Progression of Parkinson's disease as evaluated by Hoehn and Yahr stage transition times}},
volume = {25},
year = {2010}
}    

@Article{neuromorphometrics,
   Author="Bakker, R.  and Tiesinga, P.  and Kotter, R. ",
   Title="{{T}he {S}calable {B}rain {A}tlas: {I}nstant {W}eb-{B}ased {A}ccess to {P}ublic {B}rain {A}tlases and {R}elated {C}ontent}",
   Journal="Neuroinformatics",
   Year="2015",
   Volume="13",
   Pages="353--366",
}

@article{prasuhn_machine_2020,
	title = {A machine learning-based classification approach on {Parkinson}’s disease diffusion tensor imaging datasets},
	volume = {2},
	issn = {2524-3489},
	url = {https://neurolrespract.biomedcentral.com/articles/10.1186/s42466-020-00092-y},
	doi = {10.1186/s42466-020-00092-y},
	language = {en},
	number = {1},
	urldate = {2022-10-06},
	journal = {Neurological Research and Practice},
	author = {Prasuhn, J. and Heldmann, M. and Münte, T. F. and Brüggemann, N.},
	month = dec,
	year = {2020},
	pages = {46},
}


@article{sivaranjini_deep_2020,
	title = {Deep learning based diagnosis of {Parkinson}’s disease using convolutional neural network},
	volume = {79},
	issn = {1380-7501, 1573-7721},
	url = {http://link.springer.com/10.1007/s11042-019-7469-8},
	doi = {10.1007/s11042-019-7469-8},
	language = {en},
	number = {21-22},
	urldate = {2022-10-06},
	journal = {Multimedia Tools and Applications},
	author = {Sivaranjini, S. and Sujatha, C. M.},
	month = jun,
	year = {2020},
	pages = {15467--15479},
	file = {Sivaranjini and Sujatha - 2020 - Deep learning based diagnosis of Parkinson’s disea.pdf:/home/pinon/Zotero/storage/4I2E7KJD/Sivaranjini and Sujatha - 2020 - Deep learning based diagnosis of Parkinson’s disea.pdf:application/pdf},
}




@article{bergmann_mvtec_2021,
	title = {The {MVTec} {Anomaly} {Detection} {Dataset}: {A} {Comprehensive} {Real}-{World} {Dataset} for {Unsupervised} {Anomaly} {Detection}},
	volume = {129},
	issn = {0920-5691, 1573-1405},
	shorttitle = {The {MVTec} {Anomaly} {Detection} {Dataset}},
	url = {http://link.springer.com/10.1007/s11263-020-01400-4},
	doi = {10.1007/s11263-020-01400-4},
	month = apr,
	year = {2021},
	pages = {1038--1059},
	file = {Bergmann et al. - 2021 - The MVTec Anomaly Detection Dataset A Comprehensi.pdf:/home/pinon/Zotero/storage/Z5FLNI4K/Bergmann et al. - 2021 - The MVTec Anomaly Detection Dataset A Comprehensi.pdf:application/pdf},
}

@inproceedings{bergmann_mvtec_2019,
	address = {Long Beach, CA, USA},
	title = {{MVTec} {AD} — {A} {Comprehensive} {Real}-{World} {Dataset} for {Unsupervised} {Anomaly} {Detection}},
	isbn = {978-1-72813-293-8},
	url = {https://ieeexplore.ieee.org/document/8954181/},
	doi = {10.1109/CVPR.2019.00982},
	language = {en},
	urldate = {2022-10-20},
	booktitle = {2019 {IEEE}/{CVF} {Conference} on {Computer} {Vision} and {Pattern} {Recognition} ({CVPR})},
	publisher = {IEEE},
	author = {Bergmann, Paul and Fauser, Michael and Sattlegger, David and Steger, Carsten},
	month = jun,
	year = {2019},
	pages = {9584--9592},
	file = {Bergmann et al. - 2019 - MVTec AD — A Comprehensive Real-World Dataset for .pdf:/home/pinon/Zotero/storage/UEXVX36K/Bergmann et al. - 2019 - MVTec AD — A Comprehensive Real-World Dataset for .pdf:application/pdf},
}


@inproceedings{roth_towards_2022,
	address = {New Orleans, LA, USA},
	title = {Towards {Total} {Recall} in {Industrial} {Anomaly} {Detection}},
	isbn = {978-1-66546-946-3},
	url = {https://ieeexplore.ieee.org/document/9879738/},
	doi = {10.1109/CVPR52688.2022.01392},
	language = {en},
	urldate = {2022-11-02},
	booktitle = {2022 {IEEE}/{CVF} {Conference} on {Computer} {Vision} and {Pattern} {Recognition} ({CVPR})},
	publisher = {IEEE},
	author = {Roth, Karsten and Pemula, Latha and Zepeda, Joaquin and Scholkopf, Bernhard and Brox, Thomas and Gehler, Peter},
	month = jun,
	year = {2022},
	pages = {14298--14308},
	file = {Roth et al. - 2022 - Towards Total Recall in Industrial Anomaly Detecti.pdf:/home/pinon/Zotero/storage/SMW32RAH/Roth et al. - 2022 - Towards Total Recall in Industrial Anomaly Detecti.pdf:application/pdf},
}

@inproceedings{li_cutpaste_2021,
	address = {Nashville, TN, USA},
	title = {{CutPaste}: {Self}-{Supervised} {Learning} for {Anomaly} {Detection} and {Localization}},
	isbn = {978-1-66544-509-2},
	shorttitle = {{CutPaste}},
	url = {https://ieeexplore.ieee.org/document/9578875/},
	doi = {10.1109/CVPR46437.2021.00954},
	language = {en},
	urldate = {2022-11-02},
	booktitle = {2021 {IEEE}/{CVF} {Conference} on {Computer} {Vision} and {Pattern} {Recognition} ({CVPR})},
	publisher = {IEEE},
	author = {Li, Chun-Liang and Sohn, Kihyuk and Yoon, Jinsung and Pfister, Tomas},
	month = jun,
	year = {2021},
	pages = {9659--9669},
	file = {Li et al. - 2021 - CutPaste Self-Supervised Learning for Anomaly Det.pdf:/home/pinon/Zotero/storage/SFA7ARH5/Li et al. - 2021 - CutPaste Self-Supervised Learning for Anomaly Det.pdf:application/pdf},
}

@incollection{ishikawa_patch_2021,
	address = {Cham},
	title = {Patch {SVDD}: {Patch}-{Level} {SVDD} for {Anomaly} {Detection} and {Segmentation}},
	volume = {12627},
	isbn = {978-3-030-69543-9 978-3-030-69544-6},
	shorttitle = {Patch {SVDD}},
	url = {http://link.springer.com/10.1007/978-3-030-69544-6_23},
	language = {en},
	booktitle = {Computer {Vision} – {ACCV} 2020},
	publisher = {Springer International Publishing},
	author = {Yi, Jihun and Yoon, Sungroh},
	editor = {Ishikawa, Hiroshi and Liu, Cheng-Lin and Pajdla, Tomas and Shi, Jianbo},
	year = {2021},
	doi = {10.1007/978-3-030-69544-6_23},
	pages = {375--390},
	file = {Yi and Yoon - 2021 - Patch SVDD Patch-Level SVDD for Anomaly Detection.pdf:/home/pinon/Zotero/storage/4PUV73JP/Yi and Yoon - 2021 - Patch SVDD Patch-Level SVDD for Anomaly Detection.pdf:application/pdf},
}

@inproceedings{bergmann_uninformed_2020,
	address = {Seattle, WA, USA},
	title = {Uninformed {Students}: {Student}-{Teacher} {Anomaly} {Detection} {With} {Discriminative} {Latent} {Embeddings}},
	isbn = {978-1-72817-168-5},
	shorttitle = {Uninformed {Students}},
	url = {https://ieeexplore.ieee.org/document/9157778/},
	doi = {10.1109/CVPR42600.2020.00424},
	language = {en},
	urldate = {2022-11-02},
	booktitle = {2020 {IEEE}/{CVF} {Conference} on {Computer} {Vision} and {Pattern} {Recognition} ({CVPR})},
	publisher = {IEEE},
	author = {Bergmann, Paul and Fauser, Michael and Sattlegger, David and Steger, Carsten},
	month = jun,
	year = {2020},
	keywords = {Computer Science - Computer Vision and Pattern Recognition},
	pages = {4182--4191},
	file = {Bergmann et al. - 2020 - Uninformed Students Student-Teacher Anomaly Detec.pdf:/home/pinon/Zotero/storage/CLH4QHQ6/Bergmann et al. - 2020 - Uninformed Students Student-Teacher Anomaly Detec.pdf:application/pdf;Bergmann et al. - 2020 - Uninformed Students Student-Teacher Anomaly Detec.pdf:/home/pinon/Zotero/storage/ALMWAWCJ/Bergmann et al. - 2020 - Uninformed Students Student-Teacher Anomaly Detec.pdf:application/pdf},
}


@misc{yu_fastflow_2021,
	title = {{FastFlow}: {Unsupervised} {Anomaly} {Detection} and {Localization} via {2D} {Normalizing} {Flows}},
	shorttitle = {{FastFlow}},
	url = {http://arxiv.org/abs/2111.07677},
	language = {en},
	urldate = {2022-07-22},
	publisher = {arXiv},
	author = {Yu, Jiawei and Zheng, Ye and Wang, Xiang and Li, Wei and Wu, Yushuang and Zhao, Rui and Wu, Liwei},
	month = nov,
	year = {2021},
	note = {arXiv:2111.07677 [cs]},
	keywords = {Computer Science - Computer Vision and Pattern Recognition},
	file = {Yu et al. - 2021 - FastFlow Unsupervised Anomaly Detection and Local.pdf:/home/pinon/Zotero/storage/FJJUNTHI/Yu et al. - 2021 - FastFlow Unsupervised Anomaly Detection and Local.pdf:application/pdf},
}




@article{kubat_addressing_1997,
	title = {Addressing the {Curse} of {Imbalanced} {Training} {Sets}: {One}-{Sided} {Selection}},
	journal = {Proceedings of the 14th International Conference on Machine Learning},
	author = {Kubat, Miroslav and Matwin, Stan},
	year = {1997},
	file = {kubat97addressing.pdf:/home/pinon/Zotero/storage/JRL8V3NS/kubat97addressing.pdf:application/pdf},
}

@misc{msseg,
  author = {Olivier Commowick and Audrey Istace and Michaël Kain and Baptiste Laurent and Florent Leray and Mathieu Simon and Sorina Camarasu Pop and Pascal Girard and Roxana Améli and Jean-Christophe Ferré and Anne Kerbrat, Thomas Tourdias and Frédéric Cervenansky, Tristan Glatard and Jérémy Beaumont and Senan Doyle and Florence Forbes and Jesse Knight and April Khademi and Amirreza Mahbod and Chunliang Wang and Richard McKinley and Franca Wagner and John Muschelli and Elizabeth Sweeney and Eloy Roura and Xavier Lladó and Michel M. Santos and Wellington P. Santos and Abel G. Silva-Filho and Xavier Tomas-Fernandez and Hélène Urien, Isabelle Bloch and Sergi Valverde and Mariano Cabezas and Francisco Javier Vera-Olmos and Norberto Malpica and Charles Guttmann and Sandra Vukusic, Gilles Edan and Michel Dojat and Martin Styner and Simon K. Warfield and François Cotton and Christian Barillot},
  title = {Objective Evaluation of Multiple Sclerosis Lesion Segmentation using a Data Management and Processing Infrastructure. Scientific Reports 8, 13650},
  year = 2018,
  url = {http://portal.fli-iam.irisa.fr/msseg-challenge},
}

@misc{brats,
  author = {{MICCAI}},
  title = {Brain Tumor segmentation challenge },
  year = 2020,
  url = {http://braintumorsegmentation.org/},
}

@misc{isles,
  author = {{MICCAI}},
  title = {Ischemic Stoke Lesion segmentation challenge},
  year = 2018,
  url = {www.isles-challenge.org},
}

@misc{mtop,
  author = {{MICCAI}},
  title = {Mild Traumatic Brain Injury Outcome Prediction},
  year = 2016,
  url = {www.tbichallenge.wordpress.com},
}



@article{goodfellow_generative_2014,
	title = {Generative {Adversarial} {Networks}},
	journal = {arXiv:1406.2661 [cs, stat]},
	author = {Goodfellow, Ian J. and Pouget-Abadie, Jean and Mirza, Mehdi and Xu, Bing and Warde-Farley, David and Ozair, Sherjil and Courville, Aaron and Bengio, Yoshua},
	month = jun,
	year = {2014}
}




@article{Shinde2019,
author = {Shinde, Sumeet and Prasad, Shweta and Saboo, Yash and Kaushick, Rishabh and Saini, Jitender and Pal, Pramod Kumar and Ingalhalikar, Madhura},
journal = {NeuroImage: Clinical},
pages = {101748},
title = {{Predictive markers for Parkinson's disease using deep neural nets on neuromelanin sensitive MRI}},
volume = {22},
year = {2019}
}

@article{Poldrack2019,
    author = {Poldrack, Russell A. and Huckins, Grace and Varoquaux, Gael},
    title = "{Establishment of Best Practices for Evidence for Prediction: A Review}",
    pages = {534-540},
    journal = {JAMA Psychiatry},
    year = {2019}
    }
    
@Article{Xiao2015,
   Author={Xiao, Y.  and Fonov, V.  and Beriault, S.  and Al Subaie, F.  and Chakravarty, M. M.  and Sadikot, A. F.  and Pike, G. B.  and Collins, D. L. },
   Title={{{M}ulti-contrast unbiased {M}{R}{I} atlas of a {P}arkinson's disease population}},
   Journal={Int J Comput Assist Radiol Surg},
   Year={2015},
   Volume={10},
   Pages={329--341}
}

@article{Schuff2015,
author = {Schuff, Norbert and Wu, I. Wei and Buckley, Shannon and Foster, Eric D. and Coffey, Christopher S. and Gitelman, Darren R. and Mendick, Susan and Seibyl, John and Simuni, Tanya and Zhang, Yu and Jankovic, Joseph and Hunter, Christine and Tanner, Caroline M. and Rees, Linda and Factor, Stewart and Berg, Daniela and Wurster, Isabel and Gauss, Katharina and Sprenger, Fabienne and Seppi, Klaus and Poewe, Werner and Mollenhauer, Brit and Knake, Susanne and Mari, Zoltan and Mccoy, Arita and Ranola, Madelaine and Marek, Kenneth},
journal = {Movement Disorders},
pages = {1885--1892},
title = {{Diffusion imaging of nigral alterations in early Parkinson's disease with dopaminergic deficits}},
volume = {30},
year = {2015}
}

@article{Correia2020,
author = {Correia, Marta M and Rittman, Timothy and Barnes, Christopher L and Coyle-Gilchrist, Ian T and Ghosh, Boyd and Hughes, Laura E and Rowe, James B},
journal = {Brain Communications},
title = {{Towards accurate and unbiased imaging-based differentiation of Parkinson's disease, progressive supranuclear palsy and corticobasal syndrome}},
year = {2020}
}

@article{ChandolaACM2020,
author = {Chandola, Varun and Banerjee, Arindam and Kumar, Vipin},
title = {Anomaly Detection: A Survey},
year = {2009},
issue_date = {July 2009},
publisher = {Association for Computing Machinery},
address = {New York, NY, USA},
volume = {41},
number = {3},
issn = {0360-0300},
url = {https://doi.org/10.1145/1541880.1541882},
doi = {10.1145/1541880.1541882},
journal = {ACM Comput. Surv.},
month = {jul},
articleno = {15},
numpages = {58},
keywords = {outlier detection, Anomaly detection}
}

@article{ElAzami_PlosOne2016,
doi = {10.1371/journal.pone.0161498},
author = {El Azami, M. AND Hammers, A. AND Jung, J. AND Costes, N. AND Bouet, R. AND Lartizien, C.},
journal = {PLOS ONE},
publisher = {Public Library of Science},
title = {Detection of Lesions Underlying Intractable Epilepsy on T1-Weighted MRI as an Outlier Detection Problem},
year = {2016},
month = {09},
volume = {11},
url = {https://doi.org/10.1371/journal.pone.0161498},
pages = {1-21},
number = {9},
}

\end{document}